\documentclass[preprint,5pt]{elsarticle}
\usepackage[T1]{fontenc}
\usepackage[latin9]{inputenc}
\usepackage{amsmath}
\usepackage{graphicx}
\usepackage{amssymb}
\usepackage{esint}
\usepackage{amsthm}
\usepackage{dcolumn}
\usepackage{bm}
\journal{Physics Letters A}
\begin{document}
\newtheorem{conjecture}{Conjecture}\newtheorem{corollary}{Corollary}\newtheorem{theorem}{Theorem}
\newtheorem{lemma}{Lemma}\newtheorem{observation}{Observation}\newtheorem{definition}{Definition}
\newtheorem{remark}{Remark}\global\long\global\long\def\ket#1{|#1 \rangle}
 \global\long\global\long\def\bra#1{\langle#1|}
 \global\long\global\long\def\proj#1{\ket{#1}\bra{#1}}
\begin{frontmatter}

\title{The negative probabilities and information gain in weak measurements}

\author[xidian]{Xuanmin Zhu\corref{cor}}
\ead{zhuxuanmin2006@163.com}
\author[xidian]{Qun Wei}
\author[hunan]{Quanhui Liu}
\ead{quanhuiliu@gmail.com}
\author[ustc]{Shengjun Wu}
\ead{shengjun@ustc.edu.cn}

\address[xidian]{School of Science, Xidian University, Xi'an 710071, China}
\address[hunan]{School for Theoretical Physics, and Department of Applied Physics
Hunan University, Changsha 410082, China}
\address[ustc]{Hefei National Laboratory for Physical Sciences at Microscale
and Department of Modern Physics, University of Science and Technology
of China, Hefei, Anhui 230026, China}
\cortext[cor]{Corresponding author. Tel.: +86 029 88201480.}
\begin{abstract}
We study the outcomes in a general measurement with postselection, and derive upper bounds for the pointer readings in weak measurement. The probabilities inferred from weak measurements change along with the coupling strength; and the true probabilities can be obtained when the coupling is strong enough. By calculating the information gain of the measuring device about which path the particles pass through, we show that the "negative probabilities" only emerge for cases when the information gain is little due to very weak coupling between the measuring device and the particles. When the coupling strength increases, we can unambiguously determine whether a particle passes through a given path every time, hence the average shifts always represent true probabilities, and the strange "negatives probabilities" disappear.
\end{abstract}


\end{frontmatter}

\section{Introduction}
Weak measurement, a quantum measurement process with preselection and postselection, was introduced by Aharanov \emph{et al.}~\cite{aav}. In a weak measurement, the expectation value of a quantum operator can lay outside the range of the observable's eigenvalues, and this has been confirmed in the field of quantum optics~\cite{exp1}. For very weak interaction between the measuring device and the quantum system, with appropriate initial and final states, the value of the meter's reading can be much larger than that obtained in the traditional quantum measurement, this can be viewed as an amplification effect. This effect has been used to implement high-precision measurements, a tiny spin Hall effect of light has been observed by Hosten and Kwait~\cite{hall}; small transverse deflections and frequency changes of optical beams have been amplified significantly~\citep{def1}. Because of its importance in applications, there has been much work on weak measurement~\cite{w1,w2,w3,w4,w5,w6,w7,w8,w9,w10,w11,w12,w13,w14,w15,w16,w17,w18,w19,w20,w21,w22,w23,w24,w25,w26}.

Besides its usefulness in measuring small signals, weak measurement is also used extensively to analyse the foundational questions of quantum mechanics. Weak measurement provides a new perspective to the famous Hardy's paradox~\cite{hardy1}, and the predictions by Aharonov \emph{et al.}~\cite{hardy2} have  been realized in experiments~\cite{exp2}. Using the idea of weak measurement, Lundeen \emph{et al.}~\cite{Lundeen} have directly measured the transverse spatial quantum wavefunction of photons, and Kocsis \emph{et al.}~\cite{Kocsis} have observed the average trajectories of single photons in a two-slit interferometer which could not be accomplished in traditional quantum measurements. As commented by Cho~\cite{Cho}, weird weak measurements are opening new vistas in quantum physics.

In this article we study the outcomes of the pointer readings and derive the upper bounds in a weak measurement, we apply weak measurement to analyse Hardy's paradox and discuss when the "negative probabilities" (observed in~\cite{exp2}) emerge. Just as negative kinetic energy~\cite{b01} and superluminal group velocities~\cite{b02}, observable negative probabilities seem confusing. In fact, the "negative probabilities" in Hardy's gedanken experiment are not true probabilities. The "negative probabilities" just indicate that the pointer's average shifts has an opposite sign from what is expected with the presence of positive number of particles, hence the "negative probabilities" just indicate a negative effect, actually. In the literature~\cite{Zhu}, it has been obtained that the effect of signal amplification via weak measurement only exist for the cases when the interaction between the measuring device and the quantum system is very weak. Do the "negative probabilities" only exist in the case of very weak interactions, just as does the amplification effect?  How can one view the emergence of the "negative probabilities" from an information theoretical perspective? We shall discuss these questions in this article.
\section{The range of the pointer's shifts in weak measurement}
To perform a weak measurement of an observable $\mathbf{A}$, we need four steps. First, we prepare the quantum systems to be measured in the initial state $|\psi_i\rangle$. Second, let the quantum systems interact weakly with a measuring device. Third, we perform a strong measurement and select the quantum systems in the final state $|\psi_f\rangle$. Finally, we record the readings of the measuring device conditioned on successfully obtaining the final state $|\psi_f\rangle$ of the system. The weak value was introduced by Aharonov \emph{et al.}~\cite{aav}
\begin{equation}\label{eq1}
\mathbf{A}_w=\frac{\bra{\psi_f} \mathbf{A} \ket{\psi_i}}{\langle \psi_f | \psi_i \rangle}.
\end{equation}
which can be written as $\mathbf{A}_{w}=a+ib$ (with $a, b \in \mathcal{R}$). The interaction Hamiltonian is generally modeled as
\begin{equation}\label{eq2}
H=g\delta(t-t_0)\mathbf{A}\otimes p,
\end{equation}
where $g$ is the coupling strength with $g\geq 0$ and $p$ is the pointer momentum conjugate to the position coordinate $q$. We assume that $A$ is dimensionless and we use the natural unit $\hbar=1$.  Jozsa~\cite{w2} has given the final average shifts of pointer position and momentum
\begin{equation}\label{eq:jozsa}
\begin{split}
\delta q = \left\langle q \right\rangle ' - \left\langle q \right\rangle =  g a + g b \cdot \langle \lbrace p,q \rbrace \rangle  \\
\delta p = \left\langle p \right\rangle ' -\left\langle p \right\rangle  =  2 g b \cdot \mathrm{Var}_p .
\end{split}
\end{equation}
Here $\left\langle \hat{o} \right\rangle$ denotes the expectation value of an observable $\hat{o}$ of the device in its initial state, and $\left\langle \hat{o} \right\rangle '$ with a prime denotes the corresponding value in the final state of the device after the interaction and postselection. $\mathrm{Var}_p=(\Delta p)^2$ ($\mathrm{Var}_q=(\Delta q)^2$) denotes the variance of the pointer momentum (position) in the initial pointer state, and $\lbrace p,q \rbrace = pq +qp$ denotes the anti-commutator.

When one chooses appropriate initial and final states of the system such that $\langle \psi_f | \psi_i \rangle \rightarrow 0$, both the real and imaginary parts of the weak values can become arbitrarily large, and one might think that the average shifts of the pointer's position and momentum could become arbitrarily large as well, according to Eq. (\ref{eq:jozsa}).
However, in order to obtain Eq. (\ref{eq:jozsa}), approximations are used and only the first-order terms of $g$ are kept; the approximations as well as Eq. (\ref{eq:jozsa}) are no longer valid when $\langle \psi_f | \psi_i \rangle \to 0$. It was pointed out in~\cite{wu} that the average pointer shifts may have an upper bound, and this observation was also confirmed in \cite{Zhu,Susa}.
For the case when a qubit system weakly interacts with a pointer that was initially in a Gaussian state, it is shown in~\cite{Zhu} that the maximum average pointer shift $\delta q$ ($\delta p$) over all possible pre- and post-selections (PPS)  are bounded from above by the standard deviation $\Delta q$ ($\Delta p$) of the pointer variable in the initial state, i.e., $\max \{ \delta q\} \leq \Delta q$ and $\max \{ \delta p\} \leq \Delta p$. In the following, we shall show that these upper bounds still hold for the more general cases.

Wu and Li proposed a more general and precise framework of weak measurement by retaining the second-order terms of the coupling strength $g$~\cite{wu}. When the initial pointer state $\rho_d$ satisfies $\langle p \rangle=0$ and $\langle q \rangle=0$ (these conditions can be always satisfied by choosing a suitable "zero point") and the variance of $p$ is not changing with time,  the expressions of the average shifts in $q$ and $p$ are obtained as
\begin{eqnarray}\
\delta q &=& \frac{g \mathbf{Re} {\langle \mathbf{A}\rangle }_w }{1+g^2\text{Var}_p ( {\langle \mathbf{A}\rangle}_w ^{1,1} - \mathbf{Re}{\langle \mathbf{A}^2\rangle}_w)},  \label{eq4}  \\
\delta p &=& \frac{2g \mathbf{Im} {\langle \mathbf{A}\rangle }_w \text{Var}_p}{1+g^2\text{Var}_p ({\langle \mathbf{A}\rangle}_w ^{1,1} - \mathbf{Re}{\langle \mathbf{A}^2\rangle}_w)}, \label{eq5}
\end{eqnarray}
where
\begin{equation}\label{eq6}
\mathbf{\langle A\rangle }_w=\frac{\text{tr} ( \Pi_f \mathbf{A} \rho_s )}{\text{tr} ( \Pi_f \rho_s )}, {\langle \mathbf{A}\rangle}_w^{1,1}=\frac{\text{tr} ( \Pi_f \mathbf{A} \rho_s \mathbf{A} )}{\text{tr} ( \Pi_f \rho_s )},
\end{equation}
here $\rho_s$ is the initial state of the system (preselection), and $\Pi_f$ is a general postselection that could be a projection onto a final pure state or a subspace.

When the coupling strength is very weak, i.e., $g\Delta p \ll 1$, we search for the maximum shifts of the measuring device using the expressions
in Eqs. (\ref{eq4}) and (\ref{eq5}).
The absolute value of the shift $\delta q$
\begin{equation}\label{eq7}
|\delta q|\leq \frac{g |\mathbf{\langle A \rangle}_w| }{|1+(g \Delta p)^2( {\langle \mathbf{A}\rangle}_w ^{1,1}- \mathbf{Re}{\langle \mathbf{A}^2\rangle}_w)|}.
\end{equation}
If the observable $\mathbf{A}$ is a projective operator which satisfies $\mathbf{A}^2=\mathbf{A}$, we have
\begin{equation}\label{eq8}
|\delta q|\leq \frac{g |\mathbf{\langle A \rangle}_w| }{|1+(g \Delta p)^2({\langle \mathbf{A}\rangle}_w ^{1,1} - \mathbf{Re}{ \mathbf{A}_w})|}.
\end{equation}

First we prove ${\langle \mathbf{A}\rangle}_w ^{1,1} \geq |{\langle A \rangle}_w|^2$. Let $C={\langle \mathbf{A}\rangle}_w ^{1,1}-|{\langle A \rangle}_w|^2$, we have
\begin{equation}\label{eqz01}
C=\frac{\text{tr} ( \Pi_f \mathbf{A} \rho_s \mathbf{A} )\text{tr} ( \Pi_f \rho_s )-\text{tr} ( \Pi_f \mathbf{A} \rho_s )\text{tr} ( \rho_s \mathbf{A} \Pi_f )}{(\text{tr} ( \Pi_f \rho_s ))^2}.
\end{equation}
The spectral decomposition of the operators $\rho_s$ and $\Pi_f$ can be written as
\begin{equation}\label{eqz02}
\rho_s=\sum_i p_i |\psi_i\rangle \langle \psi_i|,~~\Pi_f=\sum_j q_j |\phi_j\rangle \langle \phi_j|,
\end{equation}
where $\sum_i p_i =1$, $p_i \geq 0$, and $q_j = 0 \text{ or }1$ since $\Pi_f$ is a projective operator.
The numerator of Eq. (\ref{eqz01}) is
\begin{equation}\label{eqz03}\begin{split}
F=&\left(\sum_{ij}p_i q_j |\langle \psi_i|\mathbf{A}|\phi_j \rangle|^2\right)\left(\sum_{mk}p_m q_k |\langle \psi_m|\phi_k \rangle|^2\right)\\
&-\left|\left(\sum_{ij}p_i q_j \langle \psi_i|\mathbf{A}|\phi_j \rangle \langle \phi_j|\psi_i \rangle \right)\right|^2.
\end{split}\end{equation}
We construct two vectors
\begin{equation}\label{eqz04}\begin{split}
&|a\rangle =\sum_{ij}\sqrt{p_iq_j}\langle \psi_i|\mathbf{A}|\phi_j \rangle|i,j\rangle \\
&|b\rangle =\sum_{ij}\sqrt{p_iq_j} \langle \psi_i|\phi_j \rangle |i,j\rangle,
\end{split}\end{equation}
where $\{|i,j\rangle\}$ is a orthonormal basis satisfying $\langle i,j|i',j'\rangle=\delta_{ii'}\delta_{jj'}$. So Eq. (\ref{eqz03}) can be rewritten as
\begin{equation}\label{eqz05}
F=\langle a|a\rangle \langle b|b \rangle -|\langle a|b\rangle|^2.
\end{equation}
From Schwarz inequality we have $F\geq 0$, equality holds when $|a\rangle$ is proportional to $|b\rangle$. The denominator of the Eq. (\ref{eqz01}) is positive, so we have
\begin{equation}\label{eqz06}
{\langle \mathbf{A}\rangle}_w ^{1,1} \geq |{\langle A \rangle}_w|^2.
\end{equation}
In particular, when the decomposition of PPS satisfies the conditions that all $\langle \psi_i|\mathbf{A}|\phi_j \rangle / \langle \psi_i|\phi_j \rangle$  equal to each other, equality holds. This inequality can also be found in~\cite{kofman} for the postselection states are pure states.

As $g \Delta p \leq 1$, let $K=1+(g \Delta p)^2({\langle \mathbf{A}\rangle}_w ^{1,1}- \mathbf{Re}{ \mathbf{A}_w})$, and form Eq. (\ref{eqz06}), we have
\begin{equation}\label{eq9}\begin{split}
K &\geq 1+(g \Delta p)^2(| \mathbf{A}_w| ^{2} - |\mathbf{A}_w|) \\
&\geq (1-\frac{1}{2}g \Delta p| \mathbf{A}_w| )^2+\frac{3}{4}(g \Delta p|\mathbf{A}_w|)^2\\
&\geq 0
\end{split}\end{equation}
From Eqs. ($\ref{eq8}$) and ($\ref{eq9}$), we obtain
\begin{equation}\label{eq10}
\begin{split}
|\delta q| &\leq \frac{g |\mathbf{A}_w| }{1+(g \Delta p)^2(| \mathbf{A}_w| ^{2} - | \mathbf{A}_w|)} \\
&=\frac{g}{\frac{1}{|\mathbf{A}_w|}+(g \Delta p)^2| \mathbf{A}_w|-(g \Delta p)^2}\\
&\leq \frac{g}{2g\Delta p-(g \Delta p)^2}.
\end{split}\end{equation}
As $g \Delta p\ll 1$, ignoring the $(g \Delta p)^2$ in the denominator, we get
\begin{equation}\label{eq11}
|\delta q|\leq \frac{1}{2\Delta p}.
\end{equation}
The equality holds when $|\mathbf{A}_w| \approx \frac{1}{g\Delta p} \gg 1$.
By a similar derivation, one obtains
\begin{equation}\label{eq12}
|\delta p|\leq \Delta p.
\end{equation}
These results can also be proved for the case when the observable $\mathbf{A}$ satisfies the property $\mathbf{A}^2=1$ via a similar proof.
For a general observable $A$, we don't know how to prove Eqs. ($\ref{eq11}$) and ($\ref{eq12}$), however, we can qualitatively show that they are still valid. As the coupling strength is weak, if one wants to achieve the maximum pointer shifts, one must ensure $|\mathbf{A}_w| \gg 1$. For $|\mathbf{A}_w| \gg 1$, the PPS are approximatively orthogonal (i.e., $\text{tr} ( \Pi_f \rho_s )\to 0$). For $\text{tr} ( \Pi_f \rho_s )\to 0$, one can get $|\mathbf{A}_w| ^{2} \gg |\mathbf{Re}{\langle \mathbf{A}^2\rangle}_w|$, and ignore the term $\mathbf{Re}{\langle \mathbf{A}^2\rangle}_w$. In view of Eq. (\ref{eqz06}), from Eq. (\ref{eq8}) one has
\begin{equation}\label{eq13}
|\delta q|\leq \frac{g |\mathbf{A}_w| }{|1+(g \Delta p|\mathbf{A}_w| )^2|}\leq \frac{1}{2\Delta p}.
\end{equation}
By a similar analysis, one can also show Eq. (\ref{eq12}) is valid for any operator $\mathbf{A}$. So,  the ranges of the average pointer shifts are given by
\begin{equation}\label{eq14}
-\frac{1}{2\Delta p} \leq \delta q \leq \frac{1}{2\Delta p},-\Delta p \leq \delta p \leq \Delta p.
\end{equation}
These upper bounds of the pointer shifts do not contradict the statement that the amplification factor has no upper bound (as claimed in \cite{Susa}), since the pointer states are fixed in our case.

If the initial pointer state is a Gaussian wave function, we have $\Delta q=\frac{1}{2\Delta p}$. The ranges of the pointer shifts are
\begin{equation}\label{eq15}
-\Delta q \leq \delta q \leq \Delta q,-\Delta p \leq \delta p \leq \Delta p.
\end{equation}
These ranges are in accordance with the maximum shifts given in~\cite{Zhu, kofman}. So, when the coupling strengthen is very weak, the mean pointer's shift $\delta q$ can reach any value in $[-\Delta q,\Delta q]$ by appropriate PPS.

\section{The probabilities inferred from weak measurement}

Now we discuss how the probabilities inferred from a weak measurement can be negative. For simplicity, we consider a projective operator $\mathbf{A}=|j\rangle \langle j|$, which has eigenvalues 0 and 1. We assume the interaction between the measuring device and quantum system is described by Eq. (\ref{eq2}), the mean pointer's shift $\delta q$ must be in $[0,g]$ in a general quantum measurement without postselection. The probability of obtaining the state $|j\rangle $ is inferred from the following formula
\begin{equation}\label{eqn01}
Prob\left(| j \rangle\right)=\frac{\delta p}{g} .
\end{equation}
For example, if the mean shift of the pointer $\delta q=g$, we infer that the probability of obtaining the state $|j\rangle$ is 1. In a general quantum measurement, if the coupling strength was too weak, and the shift of the pointer is not significantly greater than the uncertainty of the pointer variable in the initial state, then we can obtain negative values of pointer shifts even though eigenvalues of the projective operator are non-negative. However, the average value of the shift is non-negative in a general quantum measurement without postselection. In the scheme of weak measurement, there is a postselection process, and we can collect more pointer shifts with negative values to obtain a negative average value by postselection. For example, we can get $\delta q=-g$ by choosing particular PPS, and from Eq. (\ref{eqn01}) we infer that the probability of obtaining state $|j\rangle$ is $-1$, then a "negative probability" emerges in weak measurement.

In the following, we shall see how the "negative probabilities" emerge in the famous Hardy's paradox. In 1992, Hardy proposed a gedanken experiment which comprises two Mach-Zehnder interferometers  (see Fig. 1) to refute the possibility of Lorentz-invariant elements of reality~\cite{hardy1}. In Hardy's setup, an electron and a positron were injected simultaneity into the two interferometers denoted by $\mathbf{MZ^-}$ and $\mathbf{MZ^+}$ respectively. By analyzing the trajectories of the particles, Hardy obtained a contradiction between realistic trajectories inferred from one particle's detection and the trajectories inferred from the other one's which is usually called Hardy's paradox.

\begin{figure}
\centering \includegraphics[scale=0.4]{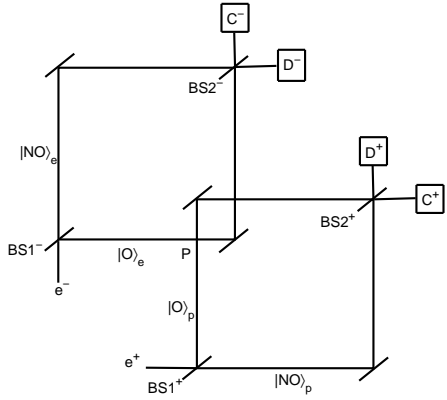} \caption{Hardy's gedanken experiment. }
\label{fig:01}
\end{figure}

The arms of the interferometers are labeled as "overlapping" $|O\rangle$, and "nonoverlapping" $|NO\rangle$, in Fig. 1. The setup is arranged such that, if an electron takes path $|O\rangle_{e}$ inside $\mathbf{MZ^-}$ and a positron takes path $|O\rangle_{p}$ inside $\mathbf{MZ^+}$, then they will meet at point \textbf{P} and annihilate each other with $100\%$ probability. The two paths $|O\rangle_e$ and $|NO\rangle_e$ in $\mathbf{MZ^-}$ can be seen as two orthonormal states for the electrons, and the two paths $|O\rangle_p$ and $|NO\rangle_p$ are two orthonormal states for the positrons. Now, we make joint weak measurement on the Hardy's gedanken experiment. The pre-selection is that no annihilation happens, so the initial state is
\begin{equation}\label{eq16}
|\psi\rangle_i=\frac{1}{\sqrt{3}}\left(|O\rangle_e|NO\rangle_p+ |NO\rangle_e |O\rangle_p+ |NO\rangle_e |NO\rangle_p\right ).
\end{equation}
The postselection corresponds to the fact that both $D^-$ and $D^+$ click, in other words, the post-selected state is
\begin{equation}\label{eq17}
|\psi\rangle_f=\frac{1}{2}\left(|O\rangle_e- |NO\rangle_e \right)\left(|O\rangle_p- |NO\rangle_p \right).
\end{equation}
The following occupation operators to be measured are projective operators
\begin{equation}\label{eq18}\begin{split}
& \mathbf{P_{O,O}}=|O\rangle_e\langle O|\otimes |O\rangle_p\langle O|, \\
&\mathbf{P_{O,NO}}=|O\rangle_e\langle O|\otimes |NO\rangle_p\langle NO|, \\
& \mathbf{P_{NO,O}}=|NO\rangle_e\langle NO|\otimes |O\rangle_p\langle O|,\\
&\mathbf{P_{NO,NO}}=|NO\rangle_e\langle NO|\otimes |NO\rangle_p\langle NO|.
\end{split}\end{equation}

We use the formalism of weak measurements derived in ~\cite{Zhu}, which is valid for arbitrary coupling strength $g$. The initial state of the measuring device is assumed to be a Gaussian wave function centered on $q=0$
\begin{equation}\label{eq19}
\Phi(q)=\frac{1}{ (2 \pi\Delta^{2})^{\frac{1}{4}}}\exp({-\frac{q^2}{4\Delta^2}}),
\end{equation}
and the standard deviations $\Delta q= \Delta$ and $\Delta p=\frac{1}{2\Delta}$. For the interaction described by Eq. (\ref{eq2}), using the Eq. (11) in~\cite{Zhu}, without any approximation, we get the mean shifts of the pointer for the four observable operators ($\mathbf{P_{O,O}}$, $\mathbf{P_{O,NO}}$, $\mathbf{P_{NO,O}}$ and $\mathbf{P_{NO,NO}}$)
\begin{equation}\label{eq22}\begin{split}
&\delta q_{O,O}=0,~~~~\delta q_{O,NO}=g,\\
&\delta q_{NO,O}=g,~~\delta q_{NO,NO}=\frac{1-2e^{-\frac{g^2}{8\Delta^2}}}{5-4e^{-\frac{g^2}{8\Delta^2}}}g.
\end{split}\end{equation}

\begin{figure}
\centering \includegraphics[scale=0.5]{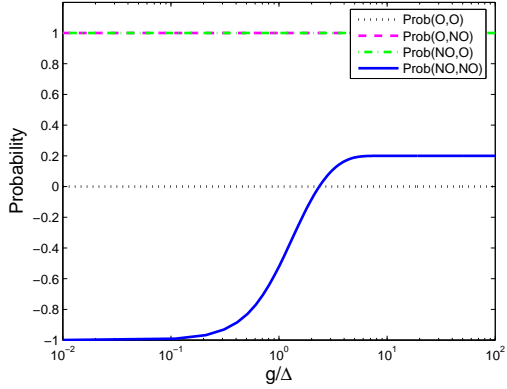} \caption{(Color online) The probabilities inferred from the mean shifts in Hardy's gedanken experiment. The probability $Prob(NO,NO)$ increases from $-1$ to ${1}/{5}$ with the coupling strength $g$.}
\label{fig:01}
\end{figure}
From Eq.(\ref{eqn01}), inferred from the mean shifts, as shown in Fig. 2, the probability of a particle passing through each path is given by
\begin{equation}\label{eq20}\begin{split}
& Prob(O,O)=0,~~~~Prob(O,NO)=1,\\
& Prob(NO,O)=1,~~Prob(NO,NO)=\frac{1-2e^{-\frac{g^2}{8\Delta^2}}}{5-4e^{-\frac{g^2}{8\Delta^2}}}.
\end{split}\end{equation}
When the interaction between the quantum system and the measuring device is weak (i.e., $g \ll  \Delta$), the probability of a particle passing through each path is given by (see Fig. 2)
\begin{equation}\label{eq21}\begin{split}
& Prob(O,O)=0,~~~~Prob(O,NO)=1,\\
& Prob(NO,O)=1,~~Prob(NO,NO)=-1.
\end{split}\end{equation}
The probability that the two particle both pass through the non-overlapping path ($|NO\rangle_e|NO\rangle_p$) is inferred to be $-1$, and the corresponding negative average pointer shift has been observed in experiments~\cite{exp2}. But from the above calculations, we see that there is no true negative probability. The observable quantities are negative average values of shifts, and the "negative probabilities" are just inferred from the shifts. So, the weird "negative probabilities" are just a negative effect, and not so hard to live with.

When the coupling strength is large (i.e., $g \gg  \Delta$), the probabilities inferred from the shifts given in Eq.(\ref{eq20}) are (see Fig. 2)
\begin{equation}\label{n02}\begin{split}
& Prob(O,O)=0,~~~~Prob(O,NO)=1,\\
& Prob(NO,O)=1,~~Prob(NO,NO)=\frac{1}{5}.
\end{split}\end{equation}
The probabilities in Eq. (\ref{n02}) are true probabilities that can also be obtained from the formulae given in~\cite{prob}, and the probability of the two particle passing through the path $|NO\rangle_e|NO\rangle_p$ is $1/5$.

From the results obtained in Eqs. (\ref{eq20}), for weak measurement, we have presented that how the probabilities inferred from the pointer shifts change along with the coupling strength. From Fig .2, it can be seen that the "pseudo negative probability" changes into true probability as the coupling strength is strong. Next, we will show the reason why this statement is valid from a perspective of the information gain of the measuring device.
\section{The negative probabilities and the information gain}
The information gain of the measuring device is the amount of information about which paths particles passing through. We use the strategy given in~\cite{zhu2} to calculate the information gain. The initial state of the measuring device is the one described by Eq. (\ref{eq19}). The information that we want to measure is about whether a electron passing through the path $|O\rangle_e$, specifically. So the observable can be written as a projector
\begin{equation}\label{eq23}
\mathbf{P_O}=|O\rangle_e\langle O|=\left(
\begin{array}{cc}
1 & 0 \\
0 & 0 \\
\end{array}
\right),
\end{equation}
in the basis $\{|O\rangle_e, |NO\rangle_e \}$. The interaction Hamiltonian is $H=g\delta(t-t_0)\mathbf{P_O}\otimes p$, by the similar calculation given in Sec. III of the literature~\cite{Zhu}, we get the maximum and minimum mean shifts of the pointer for arbitrary preselected and postselected pure states
\begin{equation}\label{eq24}\begin{split}
& \delta q_{O,\max}=\frac{g}{2}\left( 1+\frac{1}{\sqrt{1-e^{-{g^2}/{4\Delta^2}}}}  \right), \\
& \delta q_{O,\min}=\frac{g}{2}\left( 1-\frac{1}{\sqrt{1-e^{-{g^2}/{4\Delta^2}}}}  \right).
\end{split}\end{equation}
Those two extreme values are in accord with the ranges given in Eq. (\ref{eq15}). If all the average shifts of the pointer obtained in weak measurement are larger than 0, from Eq. (\ref{eqn01}), there will be no "negative probabilities" at all. If the coupling strength is weak (i.e., $g\ll \Delta$), $\delta q_{O,\min} \approx -\Delta$, we can get negative average shifts and "negative probabilities"; if the coupling strength is strong (i.e., $g\gg \Delta$), $\delta q_{O,\min} \approx 0$, all the average shifts are larger than 0, and there is no negative probability at all, as shown in Fig. 3.

\begin{figure}
\centering \includegraphics[scale=0.5]{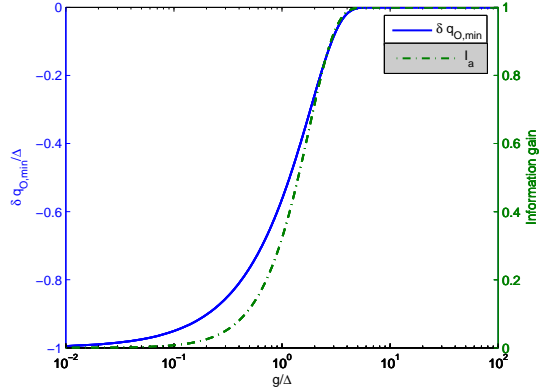} \caption{(Color online) The minimum shift of the pointer and the information gain. }
\label{fig:01}
\end{figure}

Now, we calculate the information gain of the measuring device about whether electrons passing through the path $|O\rangle_e$. It is assumed that an electron passes through the two paths $|O\rangle_e$ and $|NO\rangle_e$ with an equal probability $\frac{1}{2}$. So, the two possible states are
\begin{equation}\label{eq25}
\rho_1=\left(
\begin{array}{cc}
1 & 0 \\
0 & 0 \\
\end{array}
\right),
\rho_2=\left(
\begin{array}{cc}
0 & 0 \\
0 & 1 \\
\end{array}
\right).
\end{equation}
As defined in~\cite{zhu2}, the information gain $I_a$ of the measuring device is the mutual information of the measuring device and the quantum system which represents the correlation of the measuring device and the information source~\cite{luo}. Using the strategy given in~\cite{zhu2}, the information gain $I_a$ of the measuring device can be calculated by the following equation
\begin{equation}\label{eq26}
I_{a}=S(\rho_{R})-\frac{1}{2}\left( S(\rho_{1R})+S(\rho_{2R}) \right),
\end{equation}
where $S(\rho_{R})$ is the von Neumann entropy of $\rho_{R}$, and
\begin{equation}\label{eq27}
\rho_{R}=\frac{1}{2}\left(
\begin{array}{cc}
1 & e^{-\frac{g^2}{8\Delta^2}} \\
e^{-\frac{g^2}{8\Delta^2}} & 1 \\
\end{array}
\right),
\rho_{1R}=\left(
\begin{array}{cc}
1 & 0 \\
0 & 0 \\
\end{array}
\right),
\rho_{2R}=\left(
\begin{array}{cc}
0 & 0 \\
0 & 1 \\
\end{array}
\right),
\end{equation}
which can be obtained by Eqs. (16) and (17) in~\cite{zhu2}.
So the information gain is
\begin{equation}\label{eq28}
I_a=-\lambda\log \lambda-(1-\lambda)\log (1-\lambda),
\end{equation}
where $\lambda=(1+e^{-{g^2}/{8\Delta^2}})/2$ and the base of the logarithm function is 2, and the information gain $I_a$ is a monotonic function of coupling strength $g$.

For the weak interaction cases with $g\ll \Delta$, the information gain $I_a \approx 0$, as shown in Fig. 3. When the coupling strength is weak, since the measuring device obtains too little information, not all the shifts of the pointer could represent the correct information about which path an electron passing through. In weak measurement, we collect some shifts from all the shifts of pointer to obtain an average shift by the postselection process. If most of the shifts we collect can not represent the correct information, the probability inferred from the average shift does not represent the true probability any more. This the reason why the "negative probabilities" inferred from the negative average shift are not true probabilities when the coupling strength is two weak. For the case of strong interaction $g\gg \Delta$, $I_a \approx 1$, we can unambiguously determine whether an electron passing through the path $|O\rangle_e$ from the shift each time, and each shift can represent the correct information. Whatever shifts we choose to obtain an average shift, the probability inferred from the average is a true probability, since all the shifts represent the correct information. So, we have given the reason why not all the probabilities obtained in the case of weak coupling strength are true probabilities; while all the probabilities obtained in the case of strong coupling strength are true probabilities.

\section{conclusion}
In conclusion, we have derived the upper bounds of the pointer shifts in general weak measurements, and given the conditions for obtaining maximum average shifts. Those results could be useful for the experimentalist who try to accomplish ultra-precise measurements using weak measurements. By calculating the information gain of the measuring device, we have given the reason why the "negative probabilities" can emerge in weak measurement. We hope that our results could be useful for understanding weak measurements as well as some fundamental problems in foundations of quantum mechanics.

\section*{Acknowledgments}
This work is financially supported by the National Natural Science Foundation of China (Grants No. 11075148, and No. 11175063), and the Fundamental Research Funds
for the Central Universities.

\bibliographystyle{elsarticle-num}
\bibliography{<your-bib-database>}

\begin{thebibliography}{00}
\bibitem{aav} Y.~Aharonov, D.~Z.~Albert, and L.~Vaidman, Phys. Rev. Lett.
60 (1988) 1351.

\bibitem{exp1} N.~W.~M.~Ritchie, J.~G.~Story, and R.~G.~Hulet, Phys. Rev. Lett.
66 (1991)1107; G.~J.~Pryde, J.~L.~O'Brien, A.~G.~White, T.~C.~Ralph, and H.~M.~Wiseman, Phys. Rev. Lett.
94 (2005) 220405.

\bibitem{hall} O.~Hosten and P.~Kwiat, Science
319 (2008) 787; K.~J.~Resch, Science
319 (2008) 733.

\bibitem{def1} P.~B.~Dixon, D.~J.~Starling, A.~N.~Jordan, and J.~C.~Howell, Phys. Rev. Lett.
102 (2009) 173601; D.~J.~Starling, P.~B.~Dixon, A.~N.~Jordan, and J.~C.~Howell, Phys. Rev. A
80 (2009) 041803; D.~J.~Starling, P.~B.~Dixon, A.~N.~Jordan, and J.~C.~Howell, Phys. Rev. A
82 (2010) 063822.

\bibitem{w1} I.~M.~Duck, P.~M.~Stevenson, and E.~C.~G.~Sudarshan, Phys. Rev. D
40 (1989) 2112.

\bibitem{w2} R.~Jozsa, Phys. Rev. A
76 (2007) 044103.

\bibitem{w3} N.~Katz, M.~Neeley, M.~Ansmann, R.~C.~Bialczak, M.~Hofheinz, E.~ Lucero, A.~O'Connell,
H.~Wang,A.~N.~Cleland, J.~M.~Martinis, and A.~ N.~Korotkov, Phys. Rev. Lett.
101 (2008) 200401.

\bibitem{w4} N.~S.~Williams and A.~N.~Jordan, Phys. Rev. Lett.
100 (2008) 026804.

\bibitem{w5} V. Shpitalnik, Y. Gefen, and A. Romito, Phys. Rev. Lett.
101 (2008) 226802.

\bibitem{w6} A.~DiLorenzo and J.~C.~Egues, Phys. Rev. A
77 (2008) 042108.

\bibitem{w7} S.~Wu and K. M{\o}lmer, Phys. Lett. A
374 (2009) 34.

\bibitem{w8} N.~Brunner and C.~Simon, Phys. Rev. Lett.
105 (2010) 010405.

\bibitem{w9} Y.~Kedem and L.~Vaidman, Phys. Rev. Lett.
105 (2010) 230401.

\bibitem{w10} J.~Dressel, S.~Agarwal, and A.~N.~Jordan, Phys. Rev. Lett.
104 (2010) 240401.

\bibitem{w11} T.~Geszti, Phys. Rev. A
81 (2010) 044102.

\bibitem{w12} D.~J.~Starling, P.~B.~Dixon, N.~S.~Williams, A.~N.~Jordan, and J.~C.~Howell, Phys. Rev. A
82 (2010) 011802.

\bibitem{w13} O.~Zilberberg, A.~Romito,and Y.~Gefen, Phys. Rev. Lett.
106 (2011) 080405.

\bibitem{w14} A. Feizpour, X. Xing, and A. M. Steinberg, Phys. Rev. Lett.
107 (2011) 133603.

\bibitem{w15} N. Brunner, E. S. Polzik and C. Simon, Phys. Rev. A
84 (2011) 041804.

\bibitem{w16} S. Wu and M. \.Zukowski, Phys. Rev. Lett.
108 (2012) 080403.

\bibitem{w17} S. Pang, S. Wu, and Z. B. Chen, Phys. Rev. A
86 (2012) 022112.

\bibitem{w18} H. Kobayashi, G. Puentes, and Y. Shikano, Phys. Rev. A
86 (2012) 053805.

\bibitem{w19} J. Fischbach and M. Freyberger, Phys. Rev. A
86 (2012) 052110.

\bibitem{w20} J. Dressel and A. N. Jordan, Phys. Rev. Lett.
109 (2013) 230402.

\bibitem{w21} S. Wu, Sci. Rep. 3 (2013) 1193.

\bibitem{w22} G. C. Knee, G. A. D. Briggs, S. C. Benjamin, E. M. Gauger, Phys. Rev. A
87 (2013) 012115.

\bibitem{w23} G. Struebi, C. Bruder, Phys. Rev. Lett.
110 (2013) 083605.

\bibitem{w24} S. Tanaka, N. Yamamoto, Phys. Rev. A
88 (2013) 042116.

\bibitem{w25} X.-Y Xu, Y. Kedem, K. Sun, L. Vaidman, C.-F. Li, G.-C. Guo, Phys. Rev. Lett.
111 (2013) 033604.

\bibitem{w26} Y. Zhang, S. Wu, and Z. B. Chen, e-print arXiv: 1309.5780 [quant-ph].

\bibitem{hardy1} L.~Hardy, Phys. Rev. Lett.
68 (1992) 2981.

\bibitem{hardy2} Y.~Aharonov, A.~Botero, S.~Popescu, B.~Reznik, and J.~Tollaksen, Phys. Lett. A
301 (2002) 130.

\bibitem{exp2} J.~S.~Lundeen and A.~M.~Steinberg, Phys. Rev. Lett.
102 (2009) 020404; K. Yokota, T. Yamamoto, M. Koashi, and N. Imotoar, New J. Phys.
11 (2009) 033011.

\bibitem{b01} Y.~Aharonov, S.~Popescu, D.~Rohrlich, and L.~Vaidman, Phys. Rev. A 48 (1993) 4084.

\bibitem{b02} N.~Brunner, V.~Scarani, M.~Wegm\"{u}ller, M.~Legr\'{e},and N. Gisin, Phys. Rev. Lett. 93 (2004) 203902; D.~R.~Solli, C.~F.~McCormick, R.~Y.~Chiao, S.~Popescu, and J.~M.~Hickmann, Phys. Rev. Lett. 92 (2004) 043601.

\bibitem{Lundeen} J.~S.~Lundeen, B.~Sutherland, A.~Patel, C.~Stewart, and C. Bamber, Nature
474 (2011) 188.

\bibitem{Kocsis} S. Kocsis, B. Braverman, S. Ravets, M. J. Stevens, R. P. Mirin, L. K. Shalm, and M.Steinberg, Science
332 (2011) 1170.

\bibitem{Cho} A. Cho, Science
333 (2011) 690.

\bibitem{Zhu} X.~Zhu, Y.~Zhang, S.~Pang, C.~Qiao, Q.~Liu, and S. Wu, Phys. Rev. A
84 (2011) 052111.

\bibitem{wu} S.~Wu and Y.~Li, Phys. Rev. A
83 (2011) 052106.

\bibitem{Susa} Y. Susa, Y. Shikano, and A. Hosoya, Phys. Rev. A
85 (2012) 052110.

\bibitem{kofman} A. G. Kofman, S. Ashhab, and F. Nori, Phys. Rep. 520 (2012) 43.

\bibitem{prob} Y. Aharonov and L. Vaidman, J. Phys. A 24 (1991) 2315; Y. Aharonov and L Vaidman, e-print arXiv: 0105.101 [quant-ph].

\bibitem{zhu2} X.~Zhu, Y.~Zhang, Q.~Liu, and S. Wu, Phys. Rev. A
85 (2012) 042330.

\bibitem{luo} S. Luo, Phys. Rev. A 82 (2010) 052103.
\end{thebibliography}

\end{document}